\pdfoutput=1
\documentclass[preprint2]{aastex}

\shorttitle{Highly Polarized Microflare in S5 0716+714}
\shortauthors{Bhatta et al.}

\begin{document}

\title{Discovery of a Highly Polarized Optical Microflare in  the Blazar S5\,0716+714 During  2014 WEBT\footnote{The data collected by the WEBT Collaboration are stored in the WEBT archive; for questions regarding their availability, please contact the WEBT President Massimo Villata (\texttt{villata@oato.inaf.it).}} Campaign}

\author{G.~Bhatta\altaffilmark{1},
A.~Goyal\altaffilmark{1},
M.~Ostrowski\altaffilmark{1},
{\L}.~Stawarz\altaffilmark{1},
H.~Akitaya\altaffilmark{2},
A.~A.~Arkharov\altaffilmark{3}
R.~Bachev\altaffilmark{4},
E.~Ben\'itez\altaffilmark{5},
G.~A.~Borman\altaffilmark{6},
D.~Carosati\altaffilmark{7,8},
A.~D.~Cason\altaffilmark{9},
G.~Damljanovic\altaffilmark{10}
S.~Dhalla\altaffilmark{11},
A.~Frasca\altaffilmark{12},
D.~ Hiriart\altaffilmark{13},
S-M.~Hu\altaffilmark{14},
R.~Itoh\altaffilmark{15},
D.~Jableka\altaffilmark{1},
S.~Jorstad\altaffilmark{16,17},
K.~S.~Kawabata\altaffilmark{2},
S.~A.~Klimanov\altaffilmark{3},
O.~Kurtanidze\altaffilmark{18,19,20},
V.~M.~Larionov\altaffilmark{17,3},
D.~Laurence\altaffilmark{11},
G.~Leto\altaffilmark{12},
A.~Markowitz\altaffilmark{21},
A.~P.~Marscher\altaffilmark{16},
J.~W.~Moody\altaffilmark{22},
Y.~Moritani\altaffilmark{23},
J.~M.~Ohlert\altaffilmark{24},
A.~Di~Paola \altaffilmark{25}
C.~M.~Raiteri \altaffilmark{26}
N.~Rizzi\altaffilmark{27},
A.~C.~Sadun\altaffilmark{28},
M.~Sasada\altaffilmark{16},
S.~Sergeev\altaffilmark{6},
A.~Strigachev\altaffilmark{4},
K.~Takaki\altaffilmark{15},
I.~S.~Troitsky\altaffilmark{17},
T.~Ui\altaffilmark{15},
M.~Villata\altaffilmark{26},
O.~Vince\altaffilmark{10},
J.~R.~Webb\altaffilmark{11},
M.~Yoshida\altaffilmark{2},
and S.~Zola\altaffilmark{1,29}
}

\altaffiltext{1}{Astronomical Observatory of Jagiellonian University, ul. Orla 171, 30-244 Krakow, Poland}
\altaffiltext{2}{Hiroshima Astrophysical Science Center, Hiroshima University, Higashi-Hiroshima, Hiroshima 739-8526, Japan}
\altaffiltext{3} {Main (Pulkovo) Astronomical Observatory of RAS, Pulkovskoye shosse, 60, 196140 St. Petersburg, Russia}
\altaffiltext{4}{Institute of Astronomy, Bulgarian Academy of Sciences, 72, Tsarigradsko Shosse Blvd., 1784 Sofia Bulgaria}
\altaffiltext{5}{Instituto de Astronom\'ia, Universidad Nacional Aut\'onoma de M\'exico, Mexico DF, Mexico}
\altaffiltext{6}{Crimean Astrophysical Observatory, P/O Nauchny, Crimea, 298409, Russia}
\altaffiltext{7} {EPT Observatories, Tijarafe, La Palma, Spain}
\altaffiltext{8} {INAF, TNG Fundacion Galileo Galilei, La Palma, Spain}
\altaffiltext{9} {Private address, 105 Glen Pine Trail, Dawnsonville, GA 30534, USA}
\altaffiltext{10} {Astronomical Observatory, Volgina 7, 11060 Belgrade, Serbia}
\altaffiltext{11} {Florida International University, Miami, FL 33199, USA}
\altaffiltext{12} {Osservatorio Astrofisico di Catania, Italy}
\altaffiltext{13}{Instituto de Astronom\'ia, Universidad Nacional Aut\'onoma de M\'exico, Ensenada, Mexico}
\altaffiltext{14} {Shandong Provincial Key Laboratory of Optical Astronomy and Solar-Terrestrial Environment, Institute of Space Sciences, Shandong University at Weihai, 264209 Weihai, China}
\altaffiltext{15}{Department of Physical Science, Hiroshima University, Higashi-Hiroshima, Hiroshima 739-8526, Japan}
\altaffiltext{16} {Institute for Astrophysical Research, Boston University, 725 Commonwealth Avenue, Boston, MA 02215, USA}
\altaffiltext{17} {Astronomical Institute, St. Petersburg State University, Universitetskij Pr. 28, Petrodvorets, 198504 St. Petersburg, Russia}
\altaffiltext{18} {Abastumani Observatory, Mt. Kanobili, Abastumani, Georgia}
\altaffiltext{19} {Engelhardt Astronomical Observatory, Kazan Federal University, Tatarstan, Russia}
\altaffiltext{20} {Landessternwarte Heidelberg-Konigstuhl, Germany}
\altaffiltext{21} {University of California San Diego , 9500 Gilman Dr., La Jolla, CA 92093, USA}
\altaffiltext{22} {Physics and Astronomy Department, Brigham Young University, N283 ESC, Provo, UT, USA 84602}
\altaffiltext{23}{Kavli Institute for the Physics and Mathematics of the Universe (KavliI PMU), The University of Tokyo, 5-1-5 Kashiwa-no-Ha, Kashiwa City Chiba, 277-8583, Japan}
\altaffiltext{24} {Astronomie Stiftung Tebur, Fichtenstrasse 7, 65468 Trebur, Germany}
\altaffiltext{25}{INAF - Osservatorio Astronomico di Roma, via Frascati 33, 00040 Monte Porzio, Italy}
\altaffiltext{26}{INAF, Osservatorio Astronomico di Torino, Italy}
\altaffiltext{27} {Sirio Astronomical Observatory Castellana Grotte (Ba), Italy}
\altaffiltext{28} {Department of Physics, Univ. of Colorado Denver, CO, USA }
\altaffiltext{29}{Mt. Suhora Observatory, Pedagogical University, ul. Podchorazych 2, 30-084 Krakow, Poland}

\email{email: {\tt gopalbhatta716@gmail.com}}

\begin{abstract}
The occurrence of low-amplitude flux variations in blazars on hourly timescales, commonly known as microvariability, is still a widely debated subject in high-energy astrophysics. Several competing scenarios have been proposed to explain such occurrences, including various jet plasma instabilities leading to the formation of shocks, magnetic reconnection sites, and turbulence. In this letter we present the results of our detailed investigation of a prominent, five-hour-long optical microflare detected  during recent WEBT campaign in 2014, March 2-6 targeting the blazar 0716+714. After separating the flaring component from the underlying base emission continuum of the blazar, we find that the microflare is highly polarized, with the polarization degree $\sim (40-60)\%$\,$\pm (2-10)\%$, and the electric vector position angle $\sim (10 - 20)$\,deg\,$\pm (1-8)$\,deg slightly misaligned with respect to the position angle of the radio jet. The microflare evolution in the $(Q,\,U)$ Stokes parameter space exhibits a looping behavior with a counter-clockwise rotation, meaning polarization degree decreasing with the flux (but higher in the flux decaying phase), and approximately stable polarization angle. The overall very high polarization degree of the flare, its symmetric flux rise and decay profiles, and also its structured evolution in the $Q-U$ plane, all imply that the observed flux variation corresponds to a single emission region characterized by a highly ordered magnetic field. As discussed in the paper, a small-scale but strong shock propagating within the outflow, and compressing a disordered magnetic field component, provides a natural, though not unique, interpretation of our findings.
\end{abstract}


\section{Introduction}
\label{intro}

Blazars are known for their intense non-thermal emission and pronounced variability across the electromagnetic spectrum, resulting from the efficient energy dissipation taking place in the innermost regions of relativistic, magnetized, and non-stationary outflows --- ``jets'' --- produced by active supermassive black holes in the centers of the evolved galaxies \citep[e.g.,][]{Begelman84,Meier12}. However, our understanding of the exact physical conditions of the emitting plasma in blazar jets remains limited. The occurrence of rapid and low-amplitude flux variations on hourly timescales, commonly known as microvariability, or intra-day/night variability \citep[e.g.,][and references therein]{Wagner95}, provides additional challenge in this context, as the amount of relativistic beaming required by the plausible explanation for such phenomena is, in many cases, rather too extreme to be reconciled with the currently favored models for the jet formation in active galactic nuclei (AGN). 
  
Several competing scenarios have been proposed to explain rapid, either low- or high-amplitude variability in blazar sources. Some include {\it extrinsic causes}, such as gravitational microlensing \citep{Watson99,Webb00}; others involve {\it intrinsic origin}, including purely geometrical effects \citep[e.g., the `light house effect',][]{Camenzind92}, or various plasma instabilities leading to the formation of shocks, magnetic reconnection sites, and turbulence \citep[see the recent discussion in, e.g.,][]{Narayan12,Subramanian12,Marscher14,Sironi15,Saito15}. Since the radio-to-optical emission continuum of blazars (and BL Lacertae objects in particular) is known to be due to the synchrotron radiation by ultrarelativistic jet electrons, the temporal behavior of the optical polarization can be used as a powerful tool for diagnosing the structure of the blazar emission zone and the source of their variability. 

For example, a significant polarization degree of the synchrotron flux indicates an anisotropic distribution of the jet magnetic field, which may be related to either a large-scale uniformity of the magnetic field lines \citep[e.g.,][]{Begelman84,Lyutikov05}, or a tangled, chaotic magnetic field compressed or sheared by the flow \citep[e.g.,][]{Laing80,Laing02,Hughes85,Kollgaard90,Cawthorne90,Wardle94,Nalewajko09}. Moreover, the high duty cycle of the blazar \emph{polarization} microvariability revealed by multi-frequency optical monitoring \citep[$\gtrsim 50\%$, e.g.,][]{Andruchow05,Villforth09} indicates an origin linked to changes in the physical conditions of the jet plasma, rather than some external (to the jet) or purely geometrical effects.

In this letter we present the results of our detailed investigation of a particularly prominent and well-resolved optical microflare detected during the Whole Earth Blazar Telescope (WEBT\footnote{\texttt{http://www.to.astro.it/blazars/webt/}}) campaign targeting S5\,0716+714 in 2014, March 2-6. During the campaign, high-quality, simultaneous multi-band optical flux and polarization measurements of the source have been gathered. S5\,0716+714 is one of the brightest blazars of the ``intermediate-frequency-peaked BL Lac'' type (IBL), exhibiting persistent activity in all wavebands, including radio, optical, and $\gamma$-rays \citep[e.g.,][]{Bhatta13,Rani15}. In particular, the source is known to show rapid optical fluctuations on the timescales of minutes and hours \citep{Fan11,Bhatta13,Wu14}; high and variable optical polarization degree of about $\gtrsim 30\% $, along with fast rotations of the position angle of the electric vector, have been also observed \citep[e.g.,][]{Larionov13}. The microvariability duty cycle of S5\,0716+714 is very high \citep[$\sim 80\%-90\%$;][]{Webb07,Hu14} when compared with that of other blazars or other types of AGN \citep[see][]{Goyal13}. The selected source is therefore an ideal target to study polarization and spectra properties of the blazar optical microvariability. Below we discuss the detection of the very high polarization degree of the microflare component separated from the underlying, slowly-varying emission continuum in S5\,0716+714, indicating a highly ordered magnetic field in the sub-region of the jet responsible for the production of the observed flux enhancement.

\section{The 2014 WEBT Campaign}
\label{campaign}

The simultaneous photo-polarimetric observations of S5\,0716+714 analyzed here were obtained as a part of the WEBT campaign which lasted for five days in 2014, March 2-6. At that time the source was in a low activity-level state, increasing slowly its optical flux after the historical minimum from the end of 2013. The gathered dataset consists of high quality, high cadence multi-channel (BVRI) flux measurements from several observatories; on the two occasions, lasting for about 20\,h (56719.44--56720.27 MJD; hereafter ``Epoch~1'') and 18\,h (56721.70--56722.44 MJD; ``Epoch~2''), the high quality optical polarimetric observations were obtained  from St.Petersburg (LX-200), Crimea (AZT-8), Flagstaff, AZ (1.8\,m Perkins) and Kanata (1.5\,m) telescopes. Detailed analysis of the entire data set collected during the 2014 WEBT campaign, along with the discussion on data acquisition and reduction, will be presented elsewhere (Bhatta et al. in prep.).

\section{Analysis and Modeling Results}
\label{modeling}

Figure 1 presents the photo-polarimetric dataset gathered in the R-band for the Epoch 2 of the 2014 WEBT campaign targeting S50716+714; the flux measurements in the remaining filters are more sparse, especially in the B-band during the epoch considered. As shown, a very prominent and well-resolved microflare with the approximately symmetric and almost exponential flux rise and decay profiles --- and a sharp peak in between --- was observed in the period 79--85\,h (from the campaign starting time at 0\,h; see the dashed vertical lines in the figure). The total observed intensity of the source varied at that time by $\sim 10\%$ in $\tau_{var} \simeq 2$\,h. In order to extract the main characteristics of the flaring component, first we define the polarization degree $P\!D$, the position angle of the electric vector $\chi$, and the polarized flux $P\!F$ through the Stokes parameters $Q$ and $U$ \citep[see][]{Rybicki86},

\begin{eqnarray}
P\!D & = & \frac{\sqrt{Q^2 + U^2}}{F} \, , \nonumber \\
\chi & = & \frac{1}{2} \, \tan^{-1}\!\!\left(\frac{U}{Q}\right) \, , \quad {\rm and} \nonumber \\
P\!F & = & \sqrt{Q^2 + U^2} \, ,
\label{stokes}
\end{eqnarray}
where $F$ is the total flux of the source. 

\begin{figure}[t!]
\begin{center}
\includegraphics[angle=-90, width=1.0\columnwidth]{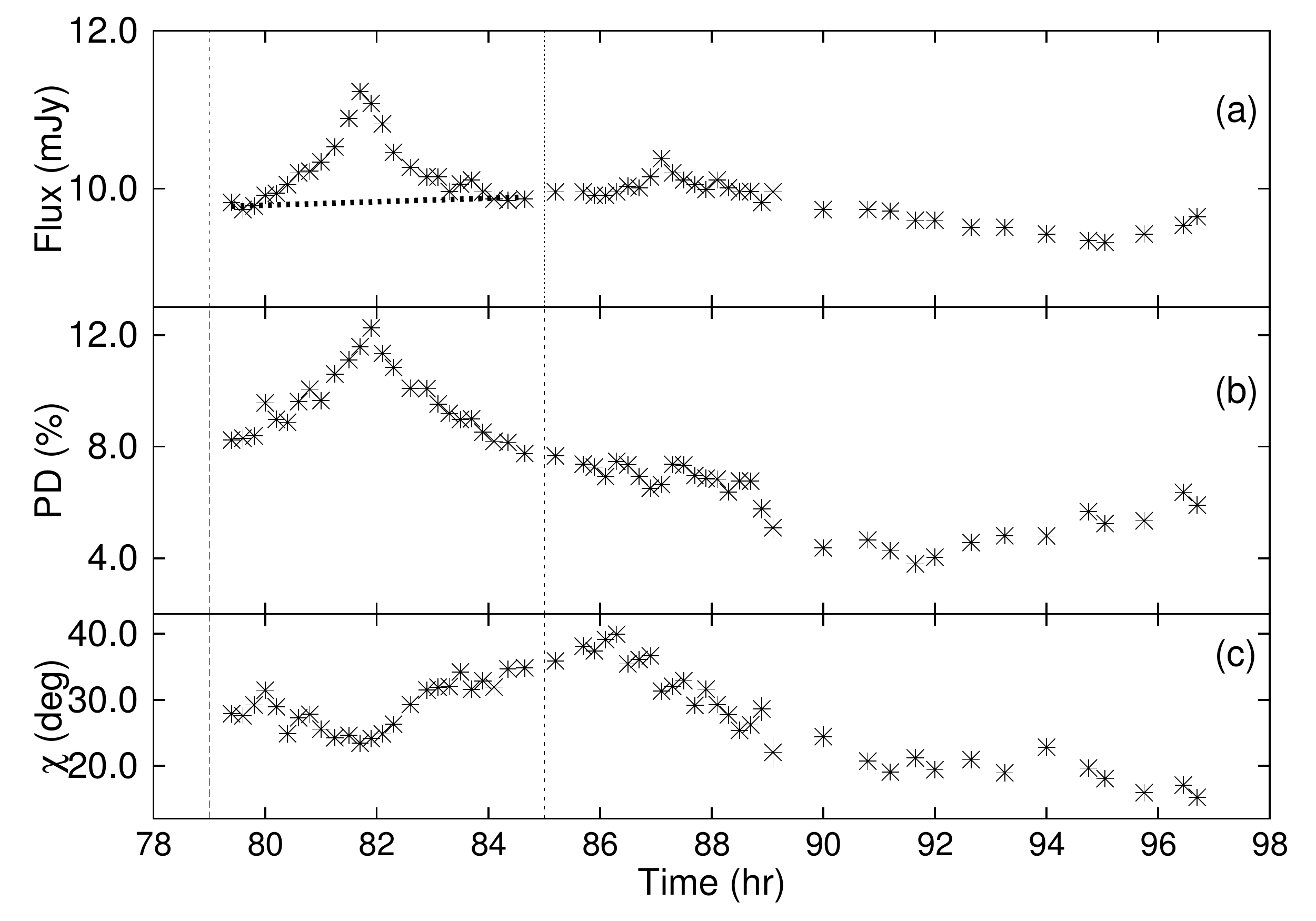}
\includegraphics[angle=-90, width=1.0\columnwidth]{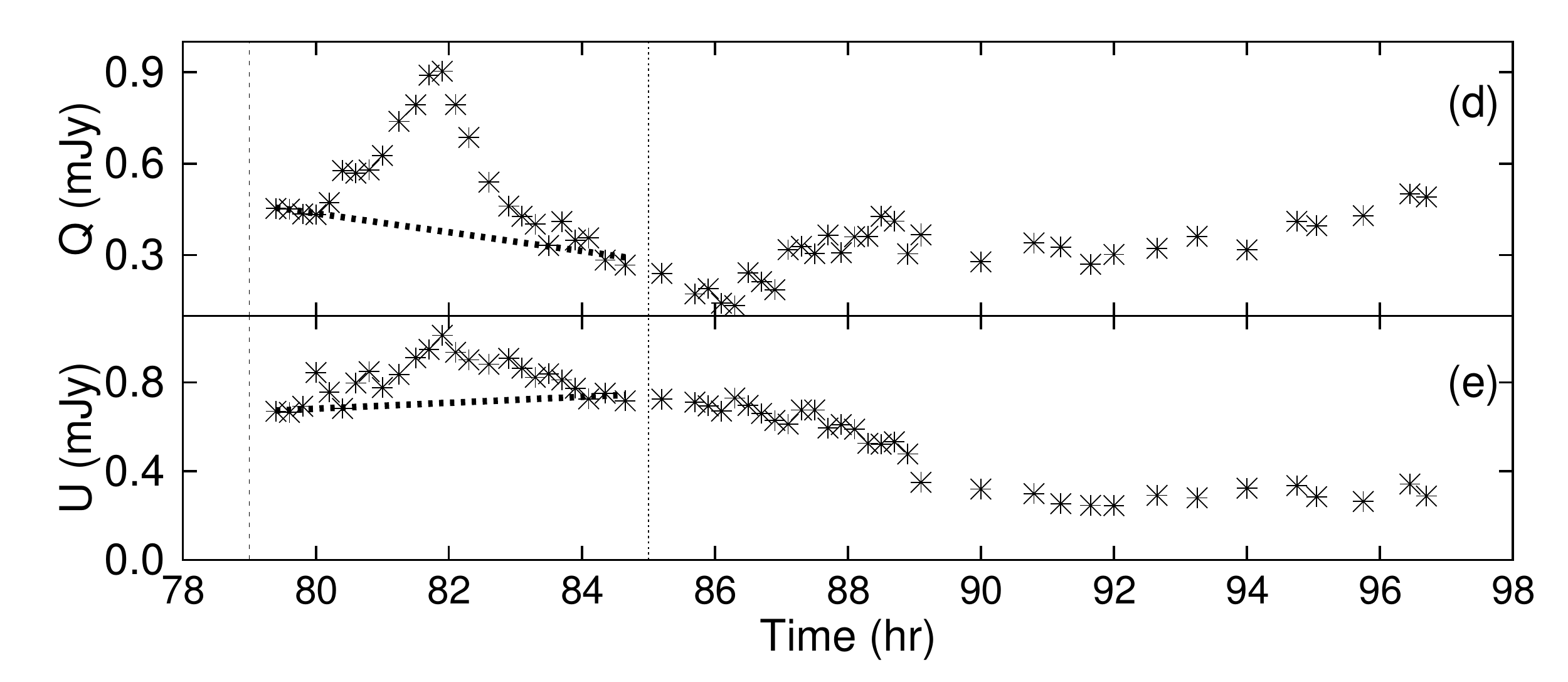}
\end{center}
\caption{Photo-polarimetric optical data collected for the blazar S5\,0716+71 during the Epoch 2 of the 2014 WEBT campaign. Subsequent panels from top to bottom present the total observed R-band flux $F$, the corresponding total polarization degree $P\!D$, the position angle of the electric vector $\chi$, and the Stokes parameters $Q$ and $U$. The dashed lines in the panels (a) and (d-e) denote the modeled ``background'' emission continuum. The period analyzed in this paper is marked by the two vertical dashed lines (between 79\,h and 85\,h after the start of the campaign).}
\label{baseflare}
\end{figure}

Next we assume that the analyzed microflare constitutes a separate emission component superimposed on the underlying, slowly variable ``background'' provided by the base emission continuum of the blazar. Due to the linearly additive properties of total flux and the Stokes $Q$ and $U$ intensities, one therefore has
\begin{eqnarray}
F & = & F_{0} + F_{1}  \, , \nonumber \\
Q & = & Q_{0} + Q_{1}  \, , \quad {\rm and} \nonumber \\
U & = & U_{0} + U_{1}  \, ,
\end{eqnarray}
where the microflare and the base emission components have been denoted by indices ``1'' and ``0'', respectively. In the modeling of this particular microflare, all the background intensities ($F_0$, $Q_0$, and $U_0$) are found from fitting the data collected just before (79-80\,h) and just after (84-85\,h) the microflare, assuming that these may change slowly (linearly) with time (see dashed lines in the corresponding panels of Figure~\ref{baseflare}). Once the base intensities are found, we subtract them from the total intensities, to obtain $F_1$, $Q_1$, and $U_1$, which further enable us to derive the basic parameters for the flaring component $P\!D_1$, $\chi_1$, and $P\!F_1$, using the standard relations given in equations (\ref{stokes}).
  
In the above analysis, we use the data exclusively in the R filter, for which continuous flux measurements are available in both the intensity and polarization for the entire duration of Epoch 2. To derive the errors for the base intensities, we take the square root of the average of the variances of the data points used for the background component fitting (i.e., just before and after the microflare). This is a reasonable assumption indeed, since in the absence of a flaring component \citep[which increases the signal to noise ratio thereby reducing the measurement uncertainty; see][]{Howell89}, the errors of the flux measurement should correspond to those of the slowly-varying background continuum. The errors in the derived quantities ($P\!D$, $\chi$, and $P\!F$), for both the base and the microflare emissions components, are derived using the standard error propagation formulae \citep{Bevington03}.

\begin{figure}[t!]
\begin{center}
\includegraphics[angle=-90, width= \columnwidth]{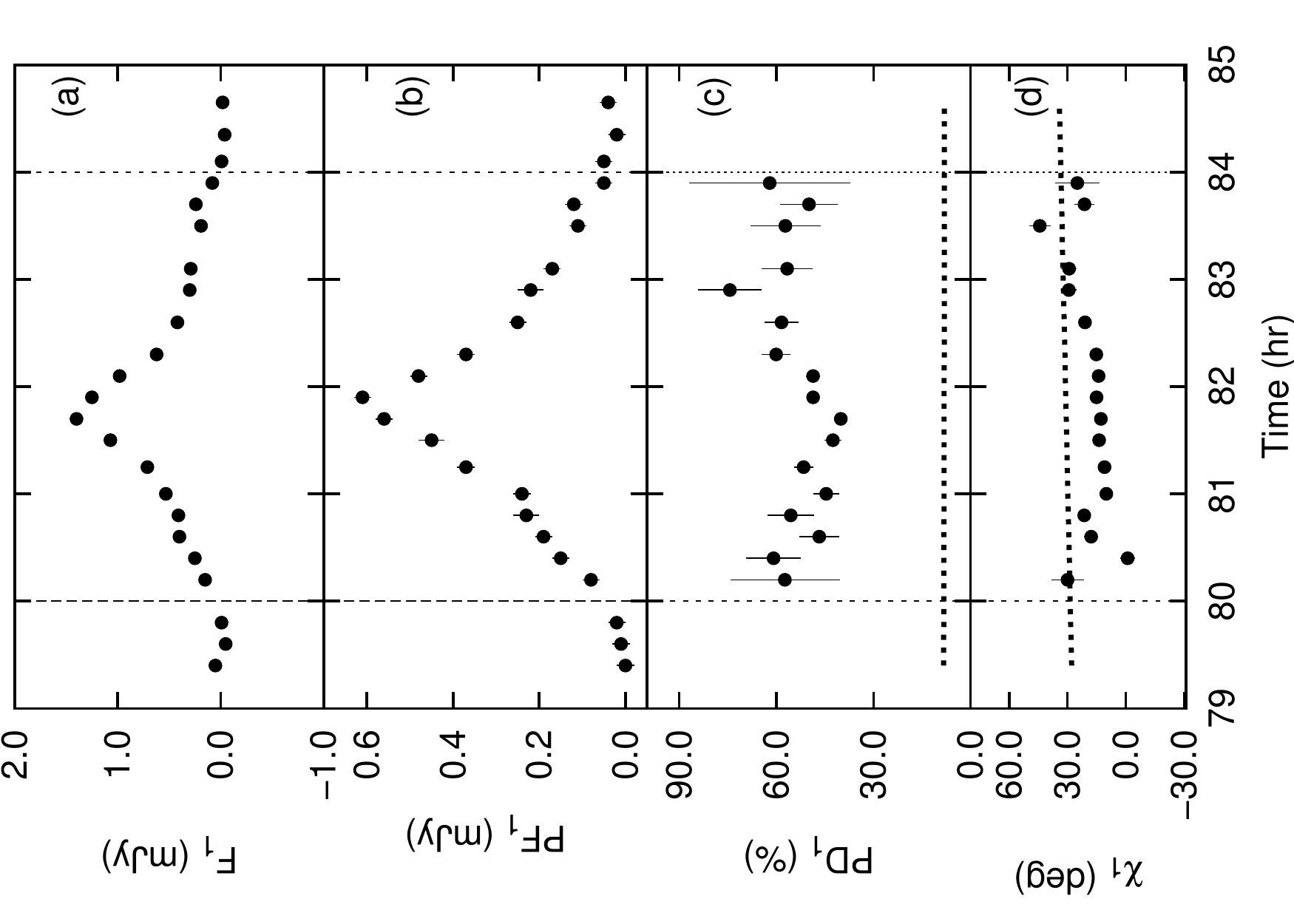}
\end{center}
\caption{The derived basic parameters of the flaring component in S5\,0716+71. Subsequent panels from top to bottom present the R-band flux $F_1$, the corresponding polarized flux $P\!F_1$, the total polarization degree $P\!D_1$, and the position angle of the electric vector $\chi_1$. Dashes in the panels (c) and (d) denote the modeled ``background'' emission continuum. The duration of the microflare is marked by the two vertical dashed lines.}
\label{shortflare}
\end{figure}

Thus derived basic parameters of the flaring component in S5\,0716+71 are presented in Figure~\ref{shortflare}. As shown, the microflare is highly polarized, with $P\!D_1 \sim (40-60)\%$\,$\pm (2-10)\%$, especially when compared with the slowly varying base emission for which $P\!D_0 \sim 8\% \pm 0.01\%$. Interestingly, the electric vector position angle of the flaring component, $\chi_1 \sim (10 - 20)$\,deg\,$\pm (1-8)$\,deg, is only slightly different from that characterizing the underlying background component, $\chi_0 \sim 30$\,deg\,$\pm 0.8$\,deg. It is important to note that, according to the high resolution radio image obtained on 2014 February 24 within the VLBA-BU-BLAZAR\footnote{\texttt{https://www.bu.edu/blazars/VLBA\_GLAST/0716.html/}} project, $\chi_0$ is close to the innermost (within 0.12\,mas from the core) position angle of the jet ($\sim$45$^\circ$), while $\chi_1$ corresponds to the position angle of the jet farther down from the core ($\sim$20$^\circ$).

\begin{figure}[t!]
\begin{center}
\includegraphics[angle=0, width=\columnwidth]{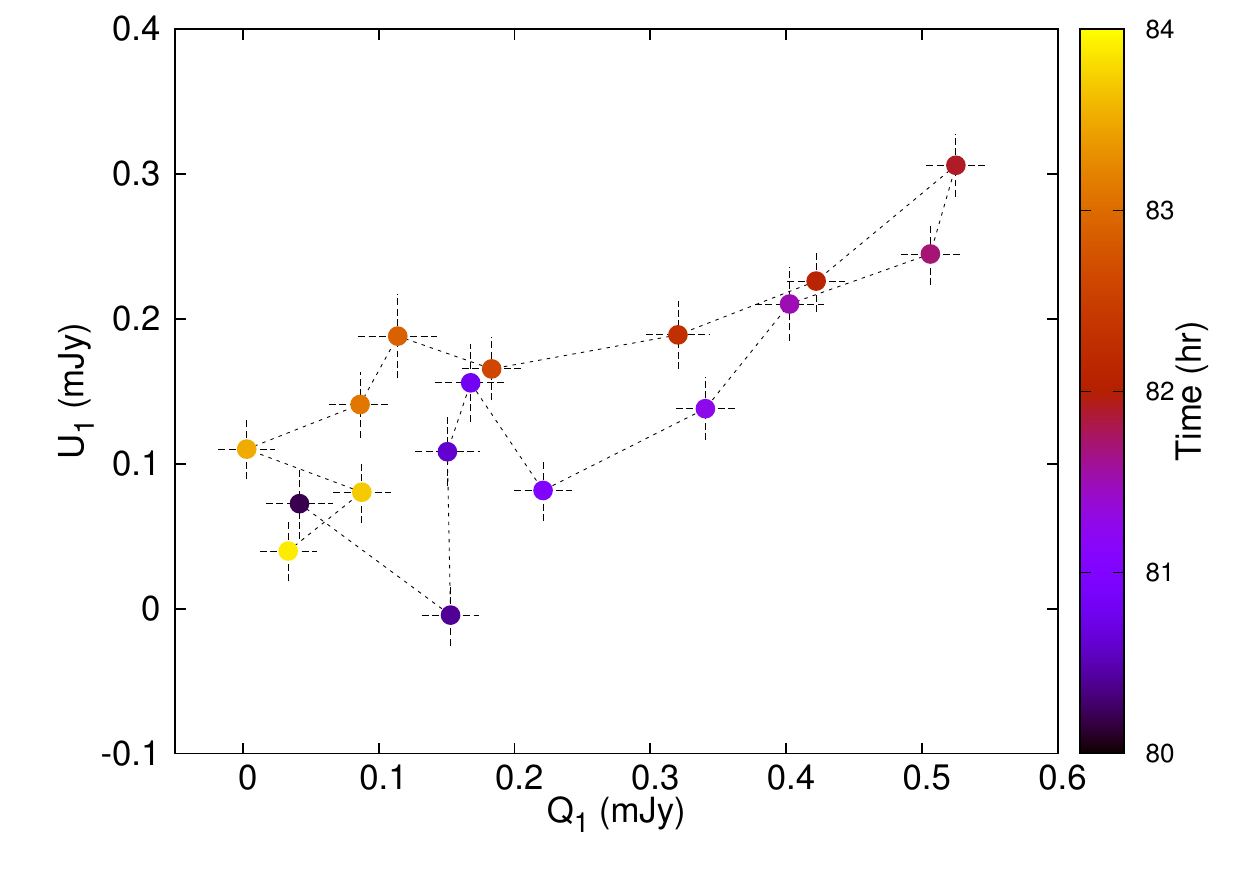}
\includegraphics[angle=0, width=\columnwidth]{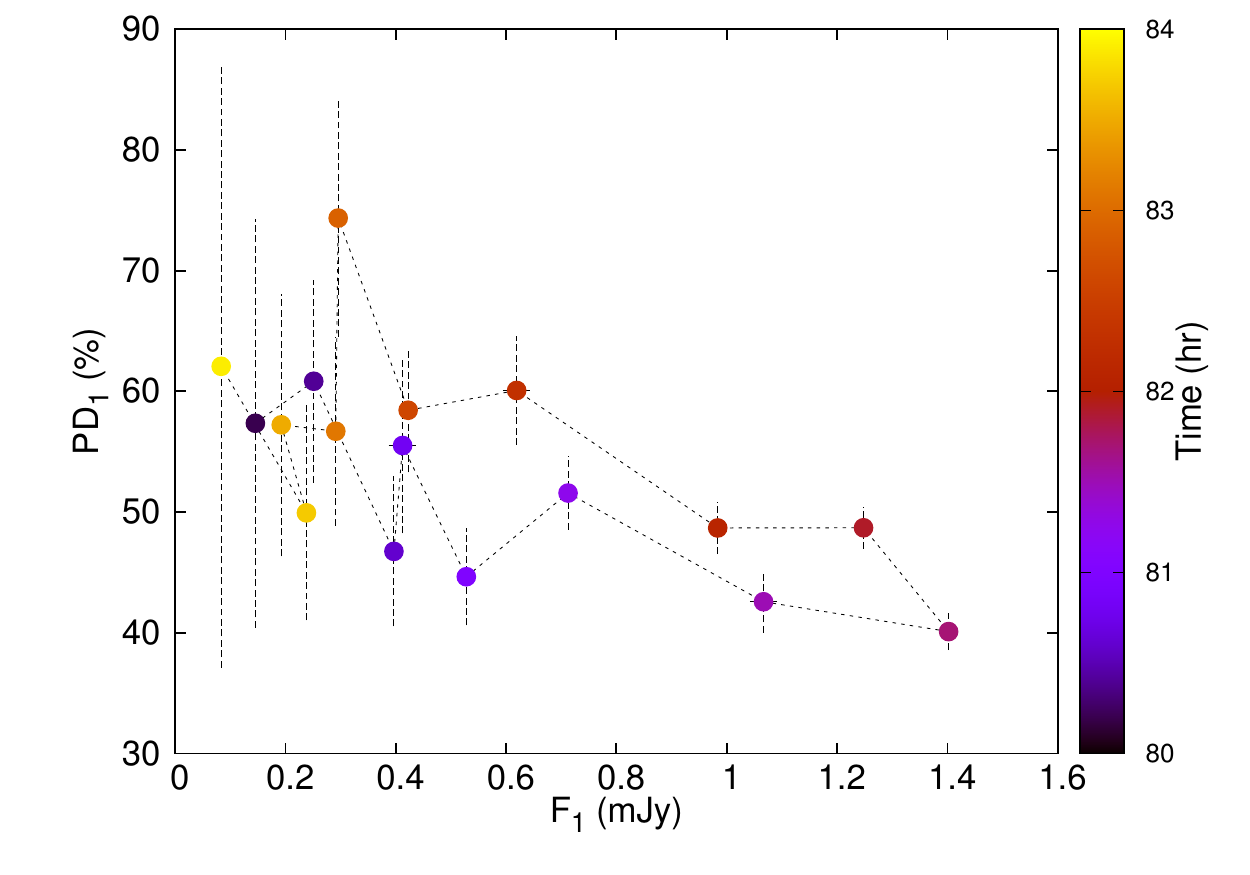}
\end{center}
\caption{The evolution of the analyzed microflare in S5\,0716+71 in the Stokes parameters plane $Q_1$ vs. $U_1$ ({\it upper panel}), and in the $ F_1-P\!D_1 $ plane ({\it lower panel}).}
\label{qu_var}
\end{figure}

In Figure~\ref{qu_var} we also present the evolution of the flaring event in the Stokes parameters plane $Q_1$ vs. $U_1$ (upper panel). As shown, there is an indication for a looping behavior with a counter-clockwise rotation, i.e. with higher polarization degree occurring in the decaying phase of the microflare. In general, the $Q-U$ diagrams illustrate the evolution of the polarized flux \emph{together} with that of the polarization angle. In the particular case presented here, the observed behavior implies therefore a consistent rate of $P\!F_1$ changes (given by the distance between the consecutive data points), with no significant rotation of $\chi_1$. This is a very interesting result as some of the similar studies on optical variability show  ``random walk''-type behavior in the $Q-U$ plane; whereas only few cases of them reveal a structured evolution \citep[e.g.,][and references therein]{Uemura10}. Moreover, during the analyzed microflare the polarization degree of the flaring component \emph{decreases} with the flux: as shown in the lower panel of Figure~\ref{qu_var}, even though the uncertainties in the polarized flux measurements are large in the begining and at the end when flaring component nearly equals the baseline emission, there is a clear counter-clockwise looping in the $P\!F_1$ vs. $F_1$ plane, with the polarization degree reaching a minimum at the level of $\sim 40\%$ at the peak of the flare. We will come back to this interesting finding in the discussion section below.

Finally, we attempt a limited spectral analysis for the microflare, using the available BVRI data, and considering the multi-band flux measurements to be simultaneous if collected within $10$\,min windows. After separating the base components of the microflare in all the filters,  we evaluate the spectral indices of the base component before and after the microflare, assuming a power-law distribution $F_{\nu} \propto \nu^{-\alpha}$. In this way we find that the spectral shape of the base emission continuum is nearly constant in time (for the period analyzed), with $\alpha_0 \simeq 1.44 \pm 0.01$. Next, we estimate the spectral index for the flaring component from the residual (total minus base) multi-band fluxes. Rather sparse BVI data allow us, however, to estimate only roughly the spectral slopes of the flaring component in the initial and final phases of the flux enhancement. The resulting spectral indices varies between $\alpha_1 \sim 1.15 \pm 0.07$ and $\sim 1.46 \pm 0.16$; large uncertainties precludes us however from making any definite statements on the spectral evolution of the flaring component.

\section{Discussion}
\label{discussion}

A few studies exist in the literature where attempts have been made to separate a base emission component from a variable emission component in the total and polarized flux optical observations of blazars. In most of the cases, the authors assumed constant background, and derived relatively high values for the polarization degree of flaring components ($\sim 20\%-50\%$), with the electric vector position angles typically aligned to the jet directions \citep{Hagen08,Sakimoto13,Morozova14,Covino15}. Particularly interesting for the work presented here are the results reported by \citet{Sasada08} for S5\,0716+714 based on the photo-polarimetric Kanata data gathered on 2007 October 20; these authors allowed for slow (linear) changes in the base emission continuum, deriving for the variable component $P\!D \approx 27\%$, with $\chi \approx 150$\,deg basically constant during the flaring event. For the flare analyzed in this paper, we found a much higher polarization degree of $\sim 40\%-60\%$, and the polarization angle $\sim 10$\,deg\,$ - 25$\,deg only slightly misaligned (by $\lesssim 20$\,deg, at most) with respect to the polarization angle of the base component, or to the position angle of the mas-scale radio jet.

The observed high polarization degree of the flare, its symmetric flux rise and decay profiles, and also its structured evolution in the $Q-U$ plane, all imply that the observed flux variation corresponds to a single and well-defined emission region characterized by a highly ordered magnetic field. A small-scale but strong shock wave propagating within the outflow, and compressing efficiently a disordered small-scale jet magnetic field component, is a natural and often invoked interpretation. And in fact, the stable electric polarization angle positioned along the jet axis can easily be reconciled with the shock interpretation. Moreover, in the case of a propagating shock, one may naturally expect a counter-clockwise looping behavior in the $F-P\!D$ plane with the polarization degree correlated with the flux, resulting from the increasing shock compression in the initial phase, followed by the decompression and spectral steepening due to radiative cooling of the emitting particles at the later stages of the shock evolution (e.g., \citealt{Hagen08}; see also in this context the discussion in \citealt{Perlman11} regarding the optical polarimetric data for the flaring HST-1 knot in the M\,87 jet). In the case of the dataset analyzed here we do observe the counter-clockwise looping, however with the polarization degree \emph{anti-correlated} with the flux. 

\begin{figure}[!t]
\begin{center}
\includegraphics[width=\columnwidth]{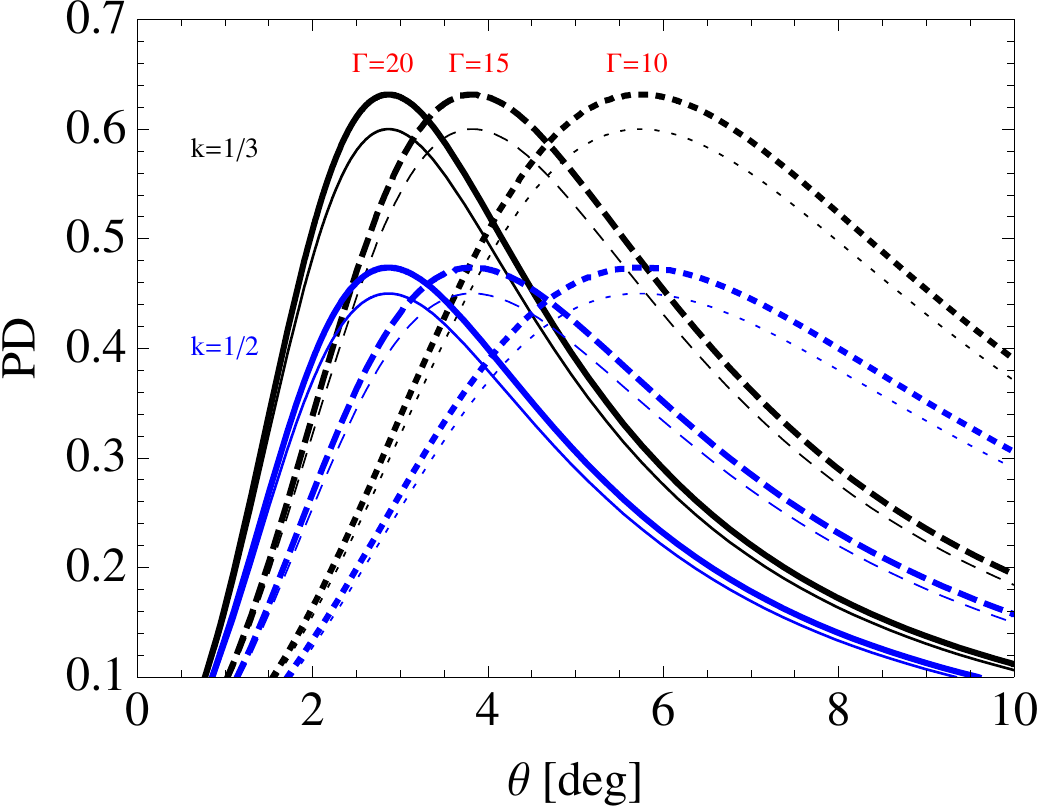}
\end{center}
\caption{The expected polarization degree $P\!D$ from a compression at the front of a strong shock as a function of the angle between the shock normal and the line of sight $\theta$, calculated using the equation (\ref{eq-shock})   for three values of the bulk Lorentz factor $\Gamma=10$, $15$, and $20$ (dotted, dashed, and solid curves, respectively) with $\alpha$ = 1.5 and 1.0 (thick and thin curves, respectively), and $k=1/3$ and $1/2$ (black and blue curves, respectively).}
\label{shock}
\end{figure}

In order to discuss this finding in more detail, though still only qualitatively, let us first note that the polarization degree expected from a shock compression can be expressed as
\begin{equation}
P\!D = \frac{3 + 3 \alpha}{5+3 \alpha} \, \frac{\delta^2 \, (1-k^2) \, \sin^2\theta}{2-\delta^2 \, (1-k^2) \, \sin^2\theta} \, ,
\label{eq-shock}
\end{equation}
where $\alpha$ is the spectral slope of the non-thermal emission continuum, $k$ is the degree of a unit length shock compression, and $\delta = 1/\Gamma \, (1-\beta \, \cos \theta)$ is the Doppler factor of the emitting plasma characterized by the bulk Lorentz factor $\Gamma = (1-\beta^2)^{-1/2}$ and the angle between the shock normal and the line of sight $\theta$ \citep[e.g.,][]{Kollgaard90}. Here we evaluate polarization degree as a funtion of $\theta$ for a number of parameters relavant to our study. Following our spectral analysis presented at the end of \S\,\ref{modeling}, we chose the two limiting values of the spectral index $\alpha = 1.0$ and 1.5, and set $k = 1/3$ as appropriate for a strong and relativistic (in the jet rest frame) shock; we also consider a weaker shocks with $k=1/2$ for comparison. We then take the three representative values of $\Gamma = 10$, 15 and 20, with the limitation of $\theta \leq 10$\,deg, for rather illustrative purposes only, noting that the analysis and modeling of the radio interferometric data regarding structural changes in the mas-scale jet of S5\,0716+71 implies the jet viewing angle $\theta_j \lesssim 5$\,deg and the jet Doppler factor $10  < \delta_j < 30$ \citep[e.g.,][and references therein]{Bach05,Rani15}. 

With the given  free parameters, we find that the expected value of the polarization degree depends strongly on the combination of $\Gamma$ and $\theta$. As illustrated in Figure~\ref{shock}, even small changes in both parameters may result in significant changes in $P\!D$ whereas the change in $\alpha$ by 0.5 may not produce appreciable change in PD. This exercise suggests in particular that the observed counter-clockwise looping behavior with the polarization degree anti-correlated with the flux could possibly be explained by assuming that at the initial stages of the shock evolution the shock normal starts to deviate from the jet axis (in a sense $\theta > \theta_j$), and that after the peak of the flare the disturbance decelerates, aligning finally its direction to that of the large-scale outflow. Interestingly, a careful look at the $\chi_1$ evolution in the bottom panel of Figure~\ref{shortflare} suggests that such a scenario may indeed be the case. Also, the maximum polarization degree expected in the model for the \emph{maximum} shock compression corresponding to $k = 1/3$, is $\simeq 60\%$, which is nicely consistent with the upper bound for $P\!D_1$ derived in \S\,\ref{modeling}. Finally, we note that in the framework of the above interpretation one may expect the other microflares in the source to be characterized by substantially lower polarization degrees and larger misalignments between the polarization vector and the jet position angle (when compared with the microflare analyzed here), due to the inevitable spread in the kinematic parameters $\Gamma$ and $\theta$ of small-scale shocks developing within the outflow. A similar geometrical approach for the interpretation of flux and $P\!D$ variability was adopted by \citet{Raiteri12} and \citet{Raiteri13} while analysing data from long-term WEBT observations, to account for the correlation between flux and $P\!D$ in the blazar 4C 38.41 and anti-correlation in BL Lacertae.

The above interpretation is obviously not a unique one, and several other models --- including for example a helical distortion in strongly magnetized plasma subjected to MHD instabilities --- may possibly account for the observational findings as well. Also, if the flaring zone consists of an underlying, highly uniform longitudinal magnetic field on top of which a shock propagates injecting freshly accelerated particles, it is plausible that the observed $F-PD$ anti-correlation is due to a reduction in the net polarization by an increasing transverse field related to the shock compression of a random, small-scale magnetic field component \citep[see in this context][]{Cawthorne93}. Only a regular, multi-band photo-polarimetric monitoring of the source with the time resolution of the sub-hour scale, allowing for a more precise characterization of multiple optical microflares in S5\,0716+714, could tighten modeling constraints and disprove various alternative scenarios, providing in this way a unique insight into the small-scale structure of relativistic outflows in AGN in general. What \emph{is} the robust conclusion from the analysis presented in this paper, however, is that the small-amplitude flux changes observed in blazar sources are related to uniform, coherent emission regions characterized by a highly ordered magnetic field and small linear sizes of the order of $\ell \sim 10^{15} \, (\tau_{var} / {\rm hr}) \, (\delta/10)$\,cm, and as such constituting (most likely) only small sub-volumes of the outflows.

\acknowledgments

The authors acknowledge support by the following grants and institutions: Polish National Science Centre grants DEC-2012/04/A/ST9/00083 (G.B., A.G., M.O., {\L}.S.) and No. 2013/09/B/ST9/ \ 00599 (S.Z.); National Natural Science Foundation of China grant 11203016 (S.-M.H.); Russian RFBR grant 15-02-00949, RFBR grant 15-32-50887 and St.Petersburg University research grant 6.38.335.2015 (St. Petersburg University team); project FR/639/6-320/12 by Shota Rustaveli National Science Foundation under contract 31/76 (Abastumani team); the observing grant support from the Institute of Astronomy and Rozhen National Astronomical Observatory, Bulgarian Academy of Sciences, and the projects No 176011, No 176004 and No 176021 supported by the Ministry of Education, Science and Technological Development of the Republic of Serbia (G.D. and O.V.); NASA under Fermi Guest Investigator grant NNX14AQ58G (BU group); UNAM-DGAPA.PAPIIT grant 111514 (E.B.). AZT-24 observations are made within an agreement between Pulkovo, Rome and Teramo observatories.

\end{document}